\documentclass[twocolumn,times]{aastex63}

\usepackage{amsmath}
\usepackage[T1]{fontenc}
\usepackage{hyperref}
\usepackage{xcolor}
\shorttitle{Faint SMBH detection via Roman}
\shortauthors{Mainak Singha~et~al.}

\graphicspath{{./}{figures/}}
\DeclareRobustCommand{\rev}[1]{#1}
\DeclareRobustCommand{\newrev}[1]{#1}
\newenvironment{revision}{}{}

\begin{document}
\title{Spectroscopic identification of $z>7$ quasars with Roman Space Telescope}

\author[0000-0001-5687-1516]{Mainak~Singha}
\affiliation{Astrophysics Science Division, NASA, Goddard Space Flight Center, Greenbelt, MD 20771, USA}
\affiliation{Department of Physics, The Catholic University of America, Washington, DC 20064, USA}
\affiliation{Center for Research and Exploration in Space Science and Technology, NASA, Goddard Space Flight Center, Greenbelt, MD 20771, USA}

\author[0000-0003-0747-1780]{Wei Leong Tee}
\affiliation{Department of Astronomy and Astrophysics, The Pennsylvania State University, 525 Davey Lab, University Park, PA 16802, USA}
\affiliation{Institute for Gravitation and the Cosmos, The Pennsylvania State University, University Park, PA 16802, USA}

\author{Sangeeta Malhotra}
\affiliation{Astrophysics Science Division, NASA, Goddard Space Flight Center, Greenbelt, MD 20771, USA}

\author{Isak Wold}
\affiliation{Astrophysics Science Division, NASA, Goddard Space Flight Center, Greenbelt, MD 20771, USA}
\affiliation{Department of Physics, The Catholic University of America, Washington, DC 20064, USA}
\affiliation{Center for Research and Exploration in Space Science and Technology, NASA, Goddard Space Flight Center, Greenbelt, MD 20771, USA}

\author{James Rhoads}
\affiliation{Astrophysics Science Division, NASA, Goddard Space Flight Center, Greenbelt, MD 20771, USA}

\author{Xiaohui Fan}
\affiliation{Steward Observatory, University of Arizona, 933 North Cherry Avenue, Tucson, AZ 85721, USA}

%\author[0000-0002-9163-8653]{JER}%Helped a lot 

%three bullet points and one paragrpahs
% take section 2.6 and reduce

\begin{abstract}
\begin{revision}
The Roman Wide Field Instrument grism will cover a large area over $1.0$--$1.93\,\mu{\rm m}$, making it possible to search directly for Ly$\alpha$ breaks in quasars beyond $z\simeq7$. We simulate quasars at $z=7.5$--10 under HLWAS-like exposures with ESpRESSO, including PSF convolution, slitless dispersion, source and background Poisson noise, read noise, and Horne optimal extraction. Three Ly$\alpha$ equivalent-width (EW) prescriptions are used to bracket the uncertainty in line strength, and the extracted spectra are compared with M/L/T dwarfs and passive red galaxies using the spectral angle mapper. The Ly$\alpha$-break completeness remains above 90\% to $J_{\rm AB}\simeq23.7$--24.1 and reaches 50\% at $J_{\rm AB}\simeq24.3$--24.7. Beyond $J_{\rm AB}\simeq25$, the result depends strongly on the adopted EW distribution. In the fixed-redshift SAM test the QSO recovery rate is $\mathrm{TPR}\ge0.982$. \newrev{None of the noisy brown-dwarf or red-galaxy test spectra is classified as a QSO. Because those test spectra are generated from the same finite parent libraries as the reference classes, this result measures the stability of the classification against noise within those libraries rather than an independently calibrated survey false-positive rate.} At the faint end, the strongest astrophysical sensitivity in the present calculation is to the high-EW tail of the Ly$\alpha$ distribution.
\end{revision}
\end{abstract}

\section{Introduction}\label{sec:intro}
One of the central questions in cosmic-dawn astrophysics is how the first
massive black holes formed and grew. Quasars at $z\gtrsim6$ show that active
supermassive black holes were already present within the first Gyr of cosmic
history, while their spectra probe the ionization state of the intergalactic
medium during reionization. The most luminous systems show strong
Gunn--Peterson absorption blueward of Ly$\alpha$, indicating substantial
neutral hydrogen at $z\sim6$--7 \citep{Fan2006,Becker2015}. Even more
strikingly, billion-solar-mass black holes have been discovered at 
$z\simeq7.5$ \citep{banados2018,Wang2021QSO}, only $\sim700$ Myr after the Big
Bang. These objects prove that rapid black-hole growth occurred early, but
they also expose a deeper limitation: the known quasars are only the rare,
luminous tail of the population.

A key unknown is the faint end. A numerous population of faint quasars
would change the inferred black-hole growth history, reshape the high-redshift
AGN luminosity function, and set the possible AGN contribution to hydrogen
reionization. An analysis at $4.5\leq z\leq6.5$ found that AGN could provide
31--75\% of ionizing photons, conditional on its adopted galaxy escape
fractions and other model assumptions \citep{Singha2025Ionizing}. Whether
that contribution persists at $z>7$ remains unmeasured. Only
$\sim70$ quasars are known at $z>6.5$, and most have
$M_{1450}<-25$ \citep{Tee2023}. At fainter magnitudes, the challenge becomes
both statistical and diagnostic. The sources are harder to detect, and the
candidate lists become increasingly contaminated. At
$J_{\rm AB}\gtrsim24$, color-selection efficiency can fall below
$\sim20\%$ because high-redshift QSOs, late-type dwarfs, and red galaxies
occupy overlapping color space once photometric scatter is included
\citep{Tee2023}. The faint-quasar problem is therefore not only a
depth problem. It is a completeness--purity problem.

Nearby compact star-forming galaxies illustrate that AGN can be difficult
to identify from a single diagnostic: X-ray and broad H$\alpha$ signatures
point to low-luminosity AGN with inferred black-hole masses of
$10^4$--$10^6\,M_\odot$ in some Green Peas \citep{Singha2025Chandra},
while five Green Peas with double-peaked narrow [O\,III] lines have
multiwavelength evidence of AGN activity \citep{Lin2023DoublePea}.
These local systems provide useful comparisons but do not directly
constrain the $z>7$ QSO luminosity function.

The first years of \textit{JWST} have made this problem more urgent. Deep
NIRCam and NIRSpec observations have uncovered a substantial population of
compact red sources --- the so-called ``Little Red Dots'' --- many with broad
H$\alpha$ emission consistent with faint AGN activity
\citep{Matthee2024,Greene2024,Kocevski2024}. JWST spectroscopy has also
identified broad-line AGN at $z>4$ with luminosities below the classical
wide-field quasar regime \citep{Harikane2023,Maiolino2023}. These sources do
not simply extend pre-\textit{JWST} quasar samples to fainter fluxes. Their
apparent abundance, compact host morphologies, broad-line widths, and inferred
black-hole masses have been argued to challenge simple extrapolations of
pre-\textit{JWST} luminosity functions and black-hole--galaxy scaling
expectations. In particular, several candidates appear overabundant relative
to conservative pre-\textit{JWST} forecasts, have black-hole masses that may
be high compared to their stellar masses, and occupy bolometric luminosities
below those probed by classical wide-field quasar searches. At still higher
redshift, GN-z11 hosts evidence for an accreting black hole at $z=10.6$
\citep{Maiolino2024}, while UHZ1 is an X-ray detected quasar candidate at
$z\simeq10.1$ whose inferred black-hole and host properties have been argued
to favor a heavy-seed origin \citep{Bogdan2024,Natarajan2024}. Together,
these discoveries suggest that early black-hole growth may extend to lower
luminosities and earlier times than classical quasar samples alone imply.
However, they come from small survey areas. \textit{JWST} can characterize
faint AGN in remarkable detail; it cannot by itself measure the abundance of
rare faint QSOs over the cosmological volumes needed for a luminosity-function
census.

\begin{revision}
Wide-area surveys already cover part of the luminosity gap between classical bright quasars and the faint AGN now seen with \textit{JWST}. SHELLQs reaches substantially fainter luminosities at $z\sim6$ \citep{Matsuoka2016SHELLQs,Matsuoka2018SHELLQs,Matsuoka2023QLF}; Euclid adds wide-area near-infrared imaging and slitless spectroscopy, while Rubin/LSST will provide deep optical dropout and variability information \citep{EuclidCollaboration2022,Ivezic2019,Yang2026Euclid}. Roman is therefore not the first survey to enter this luminosity range, and slitless spectroscopy is not unique to Roman. Its value here is the combination of survey area, angular resolution, wavelength coverage, and grism sensitivity at $z\gtrsim7.5$, where the existing wide-field samples are still sparse.
\end{revision}

\rev{Roman is well suited to extend this census to higher redshift.} The Roman Wide Field Instrument (WFI)
\rev{grism covers $1.0$--$1.93\,\mu{\rm m}$ with a point-source two-pixel resolving power $R\simeq461(\lambda/\mu{\rm m})$, or about $R\simeq460$--890 across the band \citep{Bray2024,RomanWFI2025}}, placing the Ly$\alpha$ break in band for
$z_{\rm Ly\alpha}\simeq 7.2\text{--}14.9$.
For quasars at $7.5\le z\le10$, Roman grism observations would therefore
sample the spectral feature that defines the high-redshift QSO selection
problem. Instead of relying only on red colors, slitless spectra could test the
break location, the red-side continuum shape, and the spectral curvature of
common contaminants across the observed near-infrared bandpass. This is the
key difference between a photometric candidate and a physically credible
high-redshift QSO.

In this paper, we present Roman WFI grism simulations designed to quantify the
faint $z>7.5$ QSO population that could be accessible to slitless spectroscopy
and the purity of the resulting sample under an HLWAS-like exposure model. We
simulate $z=7.5$--10 QSO spectra under three Ly$\alpha$ equivalent-width priors,
propagate them through a PSF-convolved slitless WFI grism model, add photon
counting noise from the source and sky background together with detector read
noise, extract one-dimensional spectra with Horne-style optimal extraction,
and classify detected spectra with a scale-invariant spectral-angle metric (\S~\ref{sec:sam}).
We compare against M, L, and T dwarf spectra (brown dwarf or BD) and low-redshift passive
red-galaxy (RG) templates to quantify contamination. \newrev{These simulations yield a known-redshift Roman selection function $C(J,z)$ that separates two effects:}
Ly$\alpha$-break detectability, which controls yield, and
spectral-morphology classification, which controls purity. This framework is
used to estimate how faint Roman could recover QSOs, how reliable those
detections could be, and whether Roman could access the intermediate-luminosity
population linking classical wide-field quasars to the faint AGN candidates
now emerging from \textit{JWST}.

\section{Methods}
\label{sec:methods}

We develop an end-to-end simulation and classification pipeline to assess the detectability and classification of high-redshift quasars ($z = 7.5$--$10$) with the Roman Space Telescope Wide Field Instrument (WFI) grism. The pipeline comprises four stages: (1) synthetic QSO spectral energy distribution (SED) generation, (2) grism image simulation including wavelength-dependent PSF convolution and dispersion, (3) optimal spectral extraction, and (4) spectral-angle classification against brown dwarf and red galaxy contaminants. \rev{The image simulations use isolated traces. We do not include overlap from neighboring dispersed spectra, so the extraction and classification results apply to sources for which a usable decontaminated spectrum can be obtained.}
% treated separately and not included in this analysis.

\subsection{Synthetic QSO Spectral Energy Distributions}
\label{sec:sed}

We generate a Monte Carlo population of quasar spectral energy distributions (SEDs) to forecast Ly$\alpha$ detectability in Roman as a function of apparent magnitude and redshift. This requires a population model rather than a single spectrum because QSOs vary in their UV continuum slopes and line strengths, and because attenuation through the $z\gtrsim 6$ intergalactic medium (IGM) is both strong and uncertain \citep{Fan2006,Becker2013Full,Inayoshi2020}. Our approach is to start from an empirical composite quasar template and then perturb a small set of high-leverage ingredients—continuum slope, Ly$\alpha$ equivalent width (EW), and IGM attenuation.

\begin{revision}
We use the median SDSS quasar composite of \citet{VandenBerk2001Full} as the fiducial rest-frame spectrum. It was constructed from 2204 quasars and spans approximately 800--8555\,\AA; the contributing SDSS spectra have $R\simeq1800$. The composite sets the baseline continuum and emission-line spectrum. We use \citet{Telfer2002Full} and \citet{Selsing2016Full} as supporting empirical references for the extreme-UV and UV-to-near-IR spectral shape, not as separate SAM test sets. Individual simulated quasars are generated by changing the continuum slope and Ly$\alpha$ EW, applying an IGM realization, rescaling to the target magnitude, and adding noise. The test spectra therefore differ from the fiducial composite, although they remain members of the same underlying spectral family.
\end{revision}

\begin{figure}[t]
\centering
\includegraphics[width=\columnwidth]{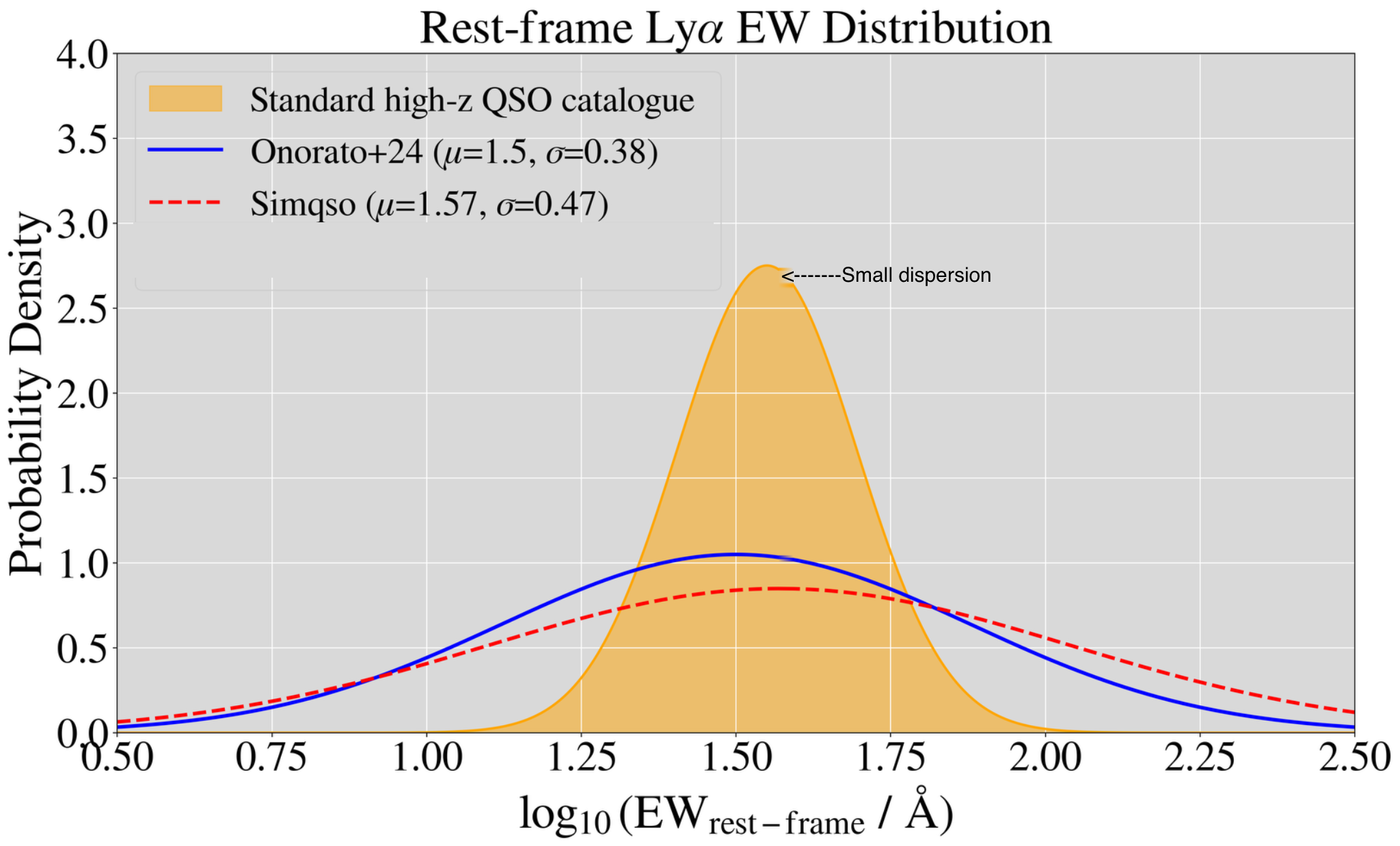}
\caption{Three Ly$\alpha$ equivalent-width priors used in the simulation
pipeline, plotted as probability densities in
$\log_{10}(\mathrm{EW}_\mathrm{rest-frame}/\text{\AA})$.
The orange filled distribution shows the narrow high-$z$ QSO catalogue
prior ($\mu=1.58$, $\sigma=0.14$, i.e.\ EW\,$\approx$\,38\,\AA{} with
$\sim30\%$ scatter), calibrated to existing $z>6$ measurements
\citep{Yang2021ApJ}.
\rev{The blue solid curve shows the Onorato-like prior \citep{Onorato2025}, with $\mu=1.50$ and $\sigma=0.38$.} The red dashed curve shows the SimQSO-like
prior \citep{SimQSO2021}, the broadest of the three ($\mu=1.57$,
$\sigma=0.47$), populating both weak-line ($\log_{10}\mathrm{EW}\sim
0.7$--$1.2$) and strong-line ($\log_{10}\mathrm{EW}\sim 1.8$--$2.3$)
tails most heavily. \rev{The three prescriptions differ in both center and width; this is not a scatter-only test. Their main practical difference at the faint end is the weight placed in the strong-line tail, where larger Ly$\alpha$ EW can move an otherwise marginal source above the $\mathrm{SNR}_{\rm break}\geq3$ threshold.}
(Section~\ref{sec:results_qso_completeness}). Running the analysis under all
three priors brackets the dominant astrophysical uncertainty in the
Roman faint-QSO forecast.}
\label{fig:ew_distribution}
\end{figure}

%The smooth transition is an assumption of population continuity designed to avoid artificial discontinuities in the EW distribution and therefore in completeness as a function of luminosity.

 %The $\tanh$ cap suppresses rare large draws without producing a hard pile-up at the cutoff, while the floor encodes the practical assumption that extremely weak Ly$\alpha$ lines contribute negligibly to the detection yield at the survey depth considered.

\begin{revision}
We model the luminosity dependence of Ly$\alpha$ EW with
\begin{equation}
\begin{aligned}
\log_{10}(\mathrm{EW}_{\rm Ly\alpha})
 &= \log_{10}(\mathrm{EW}_0) \\
 &\quad + \beta_{\rm eff}(M_{1450})
 \left(M_{1450}-M_{1450}^{\rm ref}\right)+\epsilon .
\end{aligned}
\end{equation}
where $\mathrm{EW}_0=65$~\AA\ is the reference rest-frame equivalent width, $M_{1450}^{\rm ref}=-26$ is the reference UV absolute magnitude, and $\epsilon$ is the stochastic log-EW term. \newrev{The symbol $\sigma_\epsilon$ denotes the adopted scatter of $\epsilon$ in dex; $\mu$ and $\sigma$ denote, respectively, the mean and standard deviation of the resulting $\log_{10}(\mathrm{EW}/\text{\AA})$ distribution displayed in Figure~\ref{fig:ew_distribution}. The two describe different stages of the model: $\sigma_\epsilon$ is the input scatter, $\sigma$ the dispersion of the simulated EW distribution after the luminosity dependence and sampling.} The effective Baldwin slope transitions smoothly between bright and faint regimes,
\begin{equation}
\begin{aligned}
\beta_{\rm eff}(M_{1450})
  &= (1-t)\,\beta_{\rm bright}+t\,\beta_{\rm faint},\\
 t &= \left[1+\exp\!\left(-\frac{M_{1450}-M_{\rm break}}{\Delta M}\right)\right]^{-1}.
\end{aligned}
\end{equation}
with $\beta_{\rm bright}=0.048$, $\beta_{\rm faint}=0.012$, $M_{\rm break}=-23$, and $\Delta M=0.5$ mag. We then repeat the calculation with three EW prescriptions while keeping the continuum, IGM, photometric normalization, and instrumental model fixed. \newrev{The three cases use $\sigma_\epsilon=0.15$, 0.38, and 0.47 dex, with all other ingredients identical. The narrow high-$z$ case produces $(\mu,\sigma)\simeq(1.58,0.14)$ in Figure~\ref{fig:ew_distribution}; the Onorato-like and SimQSO-like cases give $(1.50,0.38)$ and $(1.57,0.47)$.} Because both the center and width change, these should be regarded as three alternative EW distributions rather than a controlled change in scatter alone. Their median EWs are 38.0, 31.6, and 37.2~\AA, while the fraction above 100~\AA is 0.14\%, 9.4\%, and 18.0\%, respectively. We model the IGM with Lyman-series absorption plus a red-side damping wing following \citet{Fan2006,Becker2013Full,MiraldaEscude1998,MesingerHaiman2004,Santos2004,Mesinger2016Book}.
\end{revision}

Each synthetic spectrum is redshifted, attenuated by the IGM model above, and normalized to a target apparent AB magnitude in the Roman WFI F129 filter by direct filter convolution. \newrev{We refer to this as ``fixed-band normalization'': the observed-frame F129 magnitude is held at a specified value as the redshift varies, so the intrinsic UV luminosity of a source at fixed $J$ increases with $z$. We denote the F129 magnitude by $J$ below.} In-band normalization is assumed to be the most faithful mapping to observed survey photometry because the effective continuum level within a fixed observed-frame band changes with redshift as IGM absorption moves through the bandpass \citep{OkeGunn1983}. The forward SED grid uses six redshifts ($z=7.5, 8.0, 8.5, 9.0, 9.5, 10.0$) and eight apparent-magnitude bins ($J=23.0, 23.5, \ldots, 26.5$). The completeness surface in Figure~\ref{fig:lya_completeness} and the QSO SAM reference libraries use the seven bins from $J=23.0$ through 26.0; the $J=26.5$ SEDs are retained only as a faint-end check, for which the break completeness is zero across the simulated redshift grid. Within each $(J,z)$ bin, the Monte Carlo draws define the distribution of observed Ly$\alpha$ line fluxes and continua, and therefore the Ly$\alpha$ detection fraction for any specified S/N threshold. Within each $(J,z)$ bin we simulate $N_{\rm src}=50$ independent 
QSOs, each drawn with a unique combination of continuum slope 
$\alpha_\lambda$, Ly$\alpha$ EW, and IGM transmission realization 
from the distributions described above. This two-level sampling 
ensures that completeness marginalizes over both astrophysical 
variance in the source population and stochastic noise variance, 
rather than treating a single spectral shape as representative of 
the full population.

%%%%%%%%%%%%%%%%%%%%%%%%%%%%%%%%%%%%%%%%%%%%%%%%%%%%%%%%%%%%%%%%%%%%%%%%%%%%%%%%%%%%%%%%%%%%%%%%%%%%%%%%%%

\subsection{Grism Image Simulation}
\label{sec:grism_sim}

We forward-model Roman WFI slitless grism exposures using the ESpRESSO
simulation \citep{Gabrielpillai2026}. However, in this work, for each wavelength slice, we convolve the monochromatic source image with
the Roman WFI PSF evaluated at that wavelength. This step is required in
slitless spectroscopy because the PSF broadens the dispersed source profile
on the detector, affecting line contrast, source blending, and extraction
losses. We generate monochromatic PSF kernels with \texttt{STPSF}
\citep[formerly \texttt{WebbPSF};][]{Perrin2014Full} on a wavelength grid
covering the Roman grism bandpass and interpolate between kernels for
intermediate wavelengths. The convolution is
\begin{equation}
\tilde{I}_{k,s}(x,y)
=
I_{k,s}(x,y)\ast {\rm PSF}(x,y;\lambda_k),
\end{equation}
where $I_{k,s}$ is the unconvolved monochromatic image of source $s$ at
wavelength $\lambda_k$, $\tilde{I}_{k,s}$ is the PSF-convolved image, and
$\ast$ denotes two-dimensional convolution.

The interpolation assumes that the Roman WFI PSF varies smoothly across
$1.00$--$1.93\,\mu{\rm m}$ at the level required for our completeness and
extraction calculations. This is physically motivated by the diffraction
scaling $\theta\propto\lambda/D$ for an aperture diameter
$D=2.4\,{\rm m}$ \citep{Perrin2014Full}. After convolution, the wavelength
slices are dispersed through the ESpRESSO polynomial solution and weighted
by the Roman grism throughput $T(\lambda)$ to form the noiseless detector
image $I_{\rm 2D}(p,q)$ \citep{Gabrielpillai2026}, \rev{where $p$ runs along the dispersion direction and $q$ along the cross-dispersion direction. The grism effective area rises from the blue edge near $1.0\,\mu$m to a broad maximum near $1.4\,\mu$m and falls again toward $1.9\,\mu$m \citep{RomanWFI2025}. Between $z=7.5$ and 10 the Ly$\alpha$ break shifts from $1.033$ to $1.337\,\mu$m, mostly along the rising part of this response curve. The throughput therefore affects the detailed redshift dependence, but cannot by itself produce the monotonic decrease in completeness.} This noiseless image is then
passed to the noise model described in Section~\ref{sec:noise} and the
spectral extraction pipeline described in Section~\ref{sec:extraction}. The
detection statistic used to define completeness is computed from the
extracted one-dimensional spectrum and is defined in
Section~\ref{sec:detection}.

\subsection{Noise Model}
\label{sec:noise}
\begin{revision}
The simulated exposure consists of eight integrations over four visits, for a total of $2384~{\rm s}$. We include Poisson noise from the source and sky and Gaussian detector read noise. The adopted background is $0.8~e^{-}~{\rm s}^{-1}~{\rm pixel}^{-1}$, following the Roman grism simulations of \citet{Wold2024}; integrated over 2384~s this is $1907.2~e^{-}$ per pixel. We use $6~e^{-}$ per exposure for the read noise. The $0.8~e^{-}~{\rm s}^{-1}~{\rm pixel}^{-1}$ value is treated as a band-averaged background level. We do not impose a wavelength-dependent zodiacal spectrum or spatial/background variations between visits, whereas the source counts retain the wavelength dependence of the grism throughput.
\end{revision}

\subsection{Wavelength Calibration and 1D Spectral Extraction}
\label{sec:extraction}

We convert the simulated 2D grism images into one-dimensional spectra by combining a calibrated pixel-to-wavelength mapping with variance-weighted optimal extraction \citep{Horne1986}. This step closes the forward-model loop: the grism simulator predicts how photons populate detector pixels, and the extraction pipeline compresses that 2D information into a spectrum and a propagated variance model that we use to evaluate Ly$\alpha$ detectability.

\subsubsection{Wavelength calibration}
The Roman grism dispersion is approximately constant across the
$\simeq 902$-pixel trace, varying by $\lesssim 0.1$\,\AA\,pixel$^{-1}$
over the bandpass \citep{RomanWFI,Gabrielpillai2026}. We therefore adopt the approximately linear wavelength solution
\begin{equation}
\lambda(p)=\lambda_0+\delta\lambda\,p,
\label{eq:wavecal}
\end{equation}
where $\lambda_0=1000$\,nm anchors the blue end of the bandpass at
$p=0$, $p\in[0,901]$ indexes pixels along the dispersion axis, and
$\delta\lambda\simeq1.032\;\mathrm{nm\,pixel^{-1}}$. For the line-flux and variance integrations relevant to the
break-detection statistic, the residual non-linearity is well below the
5-pixel window size and we treat $\delta\lambda$ as constant.

\subsubsection{Trace initialization and centroid tracking}

We initialize each grism trace using the direct-image segmentation map and the source centroid measured at the reference wavelength of \(1.55\,\mu{\rm m}\). The segmentation-based centroid is projected onto the grism frame to define the initial trace position. Because the trace has a small cross-dispersion tilt, we track the centroid column by column using flux-weighted centroiding within a 15-pixel half-width window. The background-subtracted image is defined as
\begin{equation}
D(x,y)=N(x,y)-\widehat{B}_e ,
\end{equation}
where \(\widehat{B}_e\) is estimated empirically from source-free sidebands adjacent to the trace. The recovered background is compared against the expected value \(B_e=b\,t_{\rm tot}\), with \(b=0.8~e^-~{\rm s}^{-1}~{\rm pixel}^{-1}\) from \citet{Wold2024}. For \(t_{\rm tot}=2384~{\rm s}\), this corresponds to \(B_e\simeq1900~e^-~{\rm pixel}^{-1}\).

\begin{figure*}[t]
        \centering
        \includegraphics[width=\linewidth]{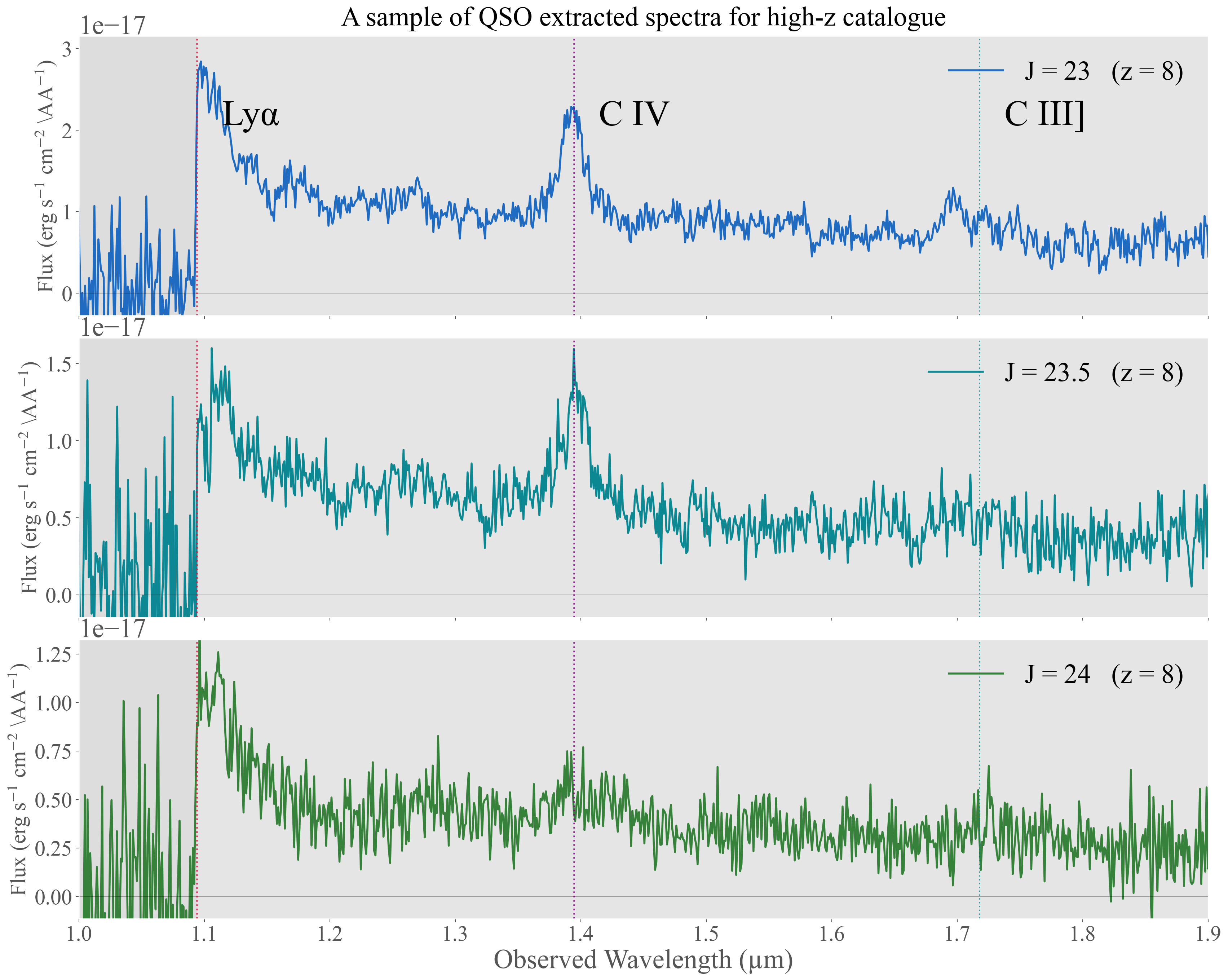}
        \caption{Extracted Roman grism spectra for a representative sample of high-redshift quasars at $z=8$, illustrating the data product underlying the completeness analysis. Panels show three sources with $J_{\rm AB}=23$, $23.5$, and $24$ (top to bottom), each plotted in a distinct color. Dotted vertical lines mark the redshifted Ly$\alpha$ ($1.094\,\mu$m), \ion{C}{4} ($1.394\,\mu$m), and \ion{C}{3}] ($1.718\,\mu$m) features, and the shaded region blueward of Ly$\alpha$ denotes the IGM-absorbed continuum. \newrev{The Ly$\alpha$ break is apparent in all three spectra. \ion{C}{4} is clear at $J_{\rm AB}=23$ and 23.5 and marginal at $J_{\rm AB}=24$, where its peak is comparable to the continuum noise.}}
        \label{fig:spectra}
    \end{figure*}

The wavelength-dependent centroid is smoothed with a Gaussian kernel of \(\sigma=3\) pixels to suppress row-to-row noise while preserving the large-scale trace tilt. We fix the centroid in columns where the positive trace flux is negligible. Explicitly tracking the tilt reduces aperture losses and prevents systematic biases in the recovered faint-source completeness. We extract one-dimensional spectra using variance-weighted optimal extraction \citep{Horne1986}. \rev{Representative extracted spectra are shown in Figure~\ref{fig:spectra}.}

\subsection{Spectral-Angle Classification}
\label{sec:sam}

At
the faint end, broad-band photometry alone does not uniquely identify
high-redshift QSOs because multiple astrophysical populations can
reproduce similar colors once noise, throughput, and extraction
losses enter. The grism spectrum provides a more discriminating
observable: high-$z$ QSOs exhibit a redshifted Ly$\alpha$ break with
a characteristic red-side continuum, whereas the dominant contaminants
— cool dwarfs and red galaxies — produce curvature and absorption
structures that generally fail to match the full-bandpass morphology
at the same wavelengths. \newrev{As in other spectroscopic classifiers, the discriminating information is the spectral shape. We quantify shape similarity with a scale-invariant angle between mean-centered spectra, so that the overall normalization and a constant background offset do not enter the comparison; wavelength-dependent extraction losses and background residuals do change the measured shape and are part of what the classifier sees.}

\subsubsection{Spectral Angle Mapper}

We use the spectral angle mapper \citep[SAM;][]{Tian2016}. SAM
compares spectra as vectors in a high-dimensional pixel space and
measures the angle between them after mean-centering, making the
similarity largely insensitive to multiplicative scaling and to small
additive offsets. This behavior is well matched to slitless grism
spectra: the absolute normalization can vary with aperture losses or
background-subtraction residuals, while classification should be
driven by the location and shape of the break and the continuum
curvature across the bandpass.

For an extracted spectrum $\mathbf{f}$ and a template $\mathbf{g}$
evaluated on the same wavelength grid of $N$ pixels, we compute the
mean-centered cosine similarity
\begin{equation}
C = \frac{\sum_{i=1}^{N} (f_i - \bar{f})(g_i - \bar{g})}
         {\sqrt{\sum_{i=1}^{N} (f_i - \bar{f})^2}
          \sqrt{\sum_{i=1}^{N} (g_i - \bar{g})^2}},
\label{eq:sam_cos}
\end{equation}
and convert it to a spectral angle $\theta_{\mathrm{SAM}} = \arccos(C)$, 
where $\bar{f}$ and $\bar{g}$ denote sample means over the evaluated
pixels. Smaller $\theta_{\mathrm{SAM}}$ indicates closer agreement in
spectral morphology. \newrev{We use uniform pixel weights in Equation~(\ref{eq:sam_cos}). The noise variance enters the extraction (Section~\ref{sec:extraction}) through the Horne profile weights; the shape comparison that follows treats every wavelength pixel equally. At the detector level the noise model of Section~\ref{sec:noise} is flat: the background and read-noise variance, $1907+288\simeq2200\,e^{-2}$ per pixel, is the same in every pixel, and a $J_{\rm AB}=24$ source adds only $\simeq30$--55~$e^{-}$ per dispersion pixel. The variance of the extracted, flux-calibrated spectrum scales as $\sigma_{\rm PSF}(\lambda)/T(\lambda)^2$, so inverse-variance weighting in flux units is, up to the slowly varying PSF-width factor, the same as evaluating the angle on the detector-count spectrum: each pixel is weighted by $T(\lambda)^2$, which favours the throughput peak near $1.4\,\mu$m over the break region at $1.0$--$1.3\,\mu$m. The two choices are the angle in flux space and the angle in count space; we adopt the former. With either fixed weighting $\theta_{\rm SAM}$ remains scale-invariant. The dominant discriminant, the Ly$\alpha$ break, is sign-definite: the blue side is zero in every QSO reference and test spectrum and non-zero in every brown-dwarf and red-galaxy template, and a positive reweighting cannot remove it, so the weighting sets the size of the $|\Delta|\simeq50^{\circ}$ margins in Figure~\ref{fig:delta_plot} rather than their sign. The archived classifier products retain the $\Delta$ values but not the per-pixel vectors, so the count-space angle is left for the calibration of the classifier on Roman data, where the measured variance array, including the wavelength-dependent zodiacal background, should set the weights.} We interpolate all templates onto the extracted wavelength grid before computing $\theta_{\mathrm{SAM}}$.

\subsubsection{Decision Rule}

A single angle does not define a class label because every spectrum
has nonzero mismatch to every template. We therefore compare proximity
to the QSO library against proximity to the contaminant libraries and
define the relative discriminant
\begin{equation}
\Delta \equiv
\min_{\mathrm{cont}}\theta_{\mathrm{SAM}}(\mathrm{spec},\mathrm{cont})
\;-\;
\min_{\mathrm{QSO}}\theta_{\mathrm{SAM}}(\mathrm{spec},\mathrm{QSO}),
\end{equation}
where ``cont'' denotes the contaminant class (BD or RG). The minima
are taken independently over the QSO and contaminant template
libraries, so $\Delta$ is the difference between the best-matching QSO
template angle and the best-matching contaminant template angle for
the same input spectrum. Three properties follow from this
construction: (i) $\Delta>0$ indicates that the closest-matching QSO
template yields a smaller spectral angle than the closest-matching
contaminant template, while $\Delta<0$ indicates the opposite, so the
sign of $\Delta$ is itself the class label; (ii) $|\Delta|$ measures
\newrev{the separation between the two best template angles, in degrees, and serves as the margin we use to quantify spectral separation;} and (iii)
$\Delta$ is invariant under any common multiplicative rescaling of
$\mathbf{f}$, because $\theta_{\mathrm{SAM}}$ is itself
scale-invariant, so flux miscalibration shifts $\Delta$ only weakly.
\newrev{The ``nearest-template boundary'' is the threshold $\Delta=0$: a test spectrum is labeled QSO if its smallest angle to a QSO template is smaller than its smallest angle to any contaminant template, and contaminant otherwise. Because $\Delta$ depends on the minimum angle in each library, library size and diversity matter: a larger library has more chances to contain a close match. The present boundary is specific to these reference sets; a survey classifier would need to be calibrated on an independent validation sample.}

\subsubsection{Template and Contaminant Spectral Libraries}

\rev{Here ``reference templates'' are the noiseless one-dimensional spectra stored in the SAM libraries, while ``test spectra'' are the noisy simulated or empirical spectra presented to the classifier.} The spectral-angle classifier operates on the full 902-pixel Roman wavelength vector after resampling; no common-wavelength mask is applied in the saved SAM implementation. The synthetic QSO spectra and empirical contaminant spectra are converted into Roman-resolution one-dimensional vectors by redshifting, spectral smoothing, wavelength resampling, and magnitude normalization. \rev{In the contaminant test, the noisy brown-dwarf and red-galaxy spectra are generated from the same finite parent sets used to build the reference libraries. A noisy spectrum can therefore be compared with its own noiseless parent, or with another spectrum derived from that parent. \newrev{We call a contaminant trial that lands on the QSO side ($\Delta>0$) a ``contaminant crossing''. Zero crossings mean that added noise did not move any of the library-derived contaminants across the boundary; we refer to this as ``in-library noise stability''. A ``leave-parent-out'' test would instead remove the parent spectrum of each noisy trial, and every template derived from it, from the reference library before classifying that trial.}}

\paragraph{QSOs.}
\begin{revision}
A separate QSO reference library is built for each EW prescription. Each library contains 2100 spectra: six redshifts, seven reference-library magnitude bins ($J=23.0$--26.0), and 50 SEDs per cell ($6\times7\times50$). After matching the test spectrum in redshift, 350 QSO references remain for a given EW prescription. The $J=26.5$ forward-model bin is not included in this reference library. The spectra are convolved to Roman resolution and resampled to the grism wavelength grid. Since SAM is insensitive to an overall multiplicative scale, the same spectral shapes can be compared across the tested magnitude range.
\end{revision}

\begin{revision}
The noisy QSO test spectra vary in continuum slope, Ly$\alpha$ EW, IGM transmission, magnitude, and noise. They are nevertheless generated from the same fiducial Vanden Berk composite used for the reference library. We therefore regard the SAM result as an in-family test. Its TPR may be optimistic for quasars with continuum shapes, line ratios, or absorption features not represented by this model.
\end{revision}

\begin{revision}
For the QSO comparison we restrict the reference set to the input redshift. This deliberately removes redshift recovery from the experiment and isolates the spectral-shape classification problem. A blind application would instead evaluate a redshift grid and account for the additional trials. The mean-centered SAM statistic does not use the absolute flux normalization, so the same reference shapes can be used across the tested $J$ range.
\end{revision}

\paragraph{Brown dwarfs.}
We use 30 M/L/T dwarf spectra from the SpeX Prism Library
\citep{Burgasser2014}, spanning spectral types M5--T8. These spectra provide
empirical cool-dwarf spectral shapes over the near-infrared wavelengths
where they can contaminate high-redshift QSO color selections. Where the
native SpeX resolution exceeds the Roman grism resolving power
($R\simeq600$), we convolve the spectra with a wavelength-dependent Gaussian kernel and resample them onto the Roman wavelength grid.

The SpeX spectra are converted into processed Roman-resolution spectra and used both to construct the brown-dwarf reference class and to generate the noisy contaminant test set. \rev{The stored classifier product contains $N_{\rm BD}=8000$ noisy brown-dwarf trials over the eight-point $J=23.0$--26.5 grid. These are repeated noisy realizations of a small set of empirical spectra, not 8000 independent brown-dwarf morphologies. Because the same parent set also supplies the reference class, we use this sample to assess noise stability rather than to estimate an independent survey FPR.}

\paragraph{Red galaxies.}
We construct the passive-galaxy contaminant set from SDSS DR16 spectra \citep{Ahumada2020} selected using catalogue/photometric red-galaxy criteria, including $u-r>2.2$; the color cut is not inferred from the limited wavelength span of an individual spectrum. The parent sample spans $z_{\rm SDSS}\simeq1.0$--2.5. A passive-galaxy 4000\,\AA\ break can mimic the observed high-$z$ QSO break when the relevant rest-frame wavelength is actually covered by the parent spectrum.

\begin{revision}
The SDSS spectra do not cover the full Roman $1.0$--$1.93\,\mu{\rm m}$ range after every redshift placement. In the saved SAM preprocessing, each processed galaxy spectrum is linearly interpolated onto the 902-pixel Roman grid; wavelengths outside its native support are set to zero, and spectra with positive coverage over less than 30\% of the grid are discarded. SAM is then evaluated on the full resampled vector. Some red-galaxy vectors therefore contain zero-padded intervals, and the 4000\,\AA\ region is empirical only when it lies inside the original spectral coverage. The missing fraction can be substantial: using the nominal 3800--9200\,\AA\ SDSS range, a $z_{\rm SDSS}=1$ parent placed at $z_{\rm RG}=2.5$ lacks about 34\% of the Roman interval, while the same parent placed at $z_{\rm RG}=1.5$ lacks about 84\%. Because the same padded shapes occur in the red-galaxy reference class, long zero intervals generally make a noisy red galaxy look more similar to its contaminant reference and less similar to a QSO spectrum, driving $\Delta$ toward more negative values. The padding therefore biases the present RG rejection in the optimistic direction. The archived aggregate outputs do not retain the per-vector coverage fractions, so we do not quote a median padded fraction that cannot be reconstructed from the stored results.

The passive-galaxy reference pool contains 200 SDSS DR16 spectra. The stored false-positive test product is a larger post-processing set: after redshift placement and the 30\% coverage cut, the SAM input directory contains 1420 processed red-galaxy vectors. One noisy realization of each processed vector is evaluated in each of the eight classifier magnitude bins, $J_{\rm AB}=23.0,23.5,\ldots,26.5$, giving $1420\times8=11360$ noisy trials. We report this post-selection factorization because it is the denominator actually used by the stored classifier output; the surviving products do not preserve a reliable parent-by-redshift-placement factorization for the 1420 vectors. As for the brown dwarfs, the contaminant reference and noisy test sets are not leave-parent-out, so the zero-crossing result is an in-library noise-stability measurement rather than an independent FPR.
\end{revision}

% ----------------------------------------------------------------------

% ----------------------------------------------------------------------

\subsection{Detecting the Ly$\alpha$ break}
\label{sec:detection}

We quantify the detectability of high-redshift QSOs by measuring,
in simulated Roman grism spectra, the statistical significance of
the Ly$\alpha$ break as a function of apparent magnitude and
redshift. \rev{Each $(J,z)$ bin contains 50 parent QSO SEDs. We generate 1000 independent pixel-level noise realizations of each noiseless grism image and repeat the background estimate, extraction, and break measurement for every realization. This gives $N_{\rm tot}=50\times1000=5\times10^4$ extracted spectra per bin. The 50,000 spectra sample measurement noise around 50 astrophysical SEDs; they are not 50,000 independent quasars.} We define completeness $C(J,z)$ as the fraction satisfying the break-detection criterion.

For each realization, we compute the expected observed-frame break
wavelength and convert it to a dispersion-pixel location $p_{\mathrm{break}}$
using the wavelength solution
(Equation~\ref{eq:wavecal}), and define two adjacent 5-pixel
windows around the known simulated break position: a red-side window immediately longward of
$p_{\mathrm{break}}$ and a blue-side window immediately shortward.
We compute the window-averaged fluxes, $\bar{f}_{\mathrm{red}} =
\frac{1}{N_{\mathrm{red}}}\sum_{p\in\mathcal{P}_{\mathrm{red}}}f(p)$, and $\bar{f}_{\mathrm{blue}} =
\frac{1}{N_{\mathrm{blue}}}\sum_{p\in\mathcal{P}_{\mathrm{blue}}}f(p)$
with $N_{\mathrm{red}}=N_{\mathrm{blue}}=5$, and propagate
uncertainties using the per-pixel variances from the
optimal-extraction pipeline, $\sigma_{\mathrm{red}}^{2} =
\frac{1}{N_{\mathrm{red}}^{2}}
\sum_{p\in\mathcal{P}_{\mathrm{red}}}\sigma^{2}(p)$ and $\sigma_{\mathrm{blue}}^{2} =
\frac{1}{N_{\mathrm{blue}}^{2}}
\sum_{p\in\mathcal{P}_{\mathrm{blue}}}\sigma^{2}(p)$. 
We then define
\begin{equation}
\mathrm{SNR}_{\mathrm{break}} =
\frac{\bar{f}_{\mathrm{red}}-\bar{f}_{\mathrm{blue}}}
     {\sqrt{\sigma_{\mathrm{red}}^{2}+\sigma_{\mathrm{blue}}^{2}}}.
\label{eq:snr_break}
\end{equation}
This statistic isolates the physical signature of interest: the
contrast between transmitted red-side continuum (and any surviving
Ly$\alpha$ flux) and the heavily absorbed blue side, evaluated
with the same noise propagation used in extraction.

\begin{revision}
We count a realization as a break detection when $\mathrm{SNR}_{\mathrm{break}}\ge3$ and $\bar f_{\rm red}>0$. The $3\sigma$ cut is used to generate candidates, not to claim a secure blind identification. At a specified break position, Gaussian noise gives a one-sided local false-alarm probability of $1.35\times10^{-3}$. In our simulations the break position is fixed by the input redshift. Across $z=7.5$--10, Ly$\alpha$ spans about 295 dispersion pixels; roughly 30 non-overlapping ten-pixel placements fit across that interval. Treating those placements as independent gives a simple trials estimate of $1-(1-1.35\times10^{-3})^{30}\simeq0.04$. The actual trials are correlated, so 4\% is only a scale estimate. A local discontinuity is also not unique to Ly$\alpha$: lower-redshift breaks and emission-line structure can produce similar features. \newrev{The requirement $\bar f_{\rm red}>0$ removes negative-flux artifacts from background over-subtraction; it does little against noise-driven steps. For a zero-flux spectrum with independent Gaussian windows, a step that already satisfies $\mathrm{SNR}_{\rm break}\ge3$ has $\bar f_{\rm red}>0$ with probability $1-[1-\Phi(3)]/2=0.99933$, so the sign condition rejects only 0.07\% of such false steps. The 4\% trials estimate above is therefore the relevant false-alarm scale, and confirmation rests on the photometric context and the morphology of the full spectrum.}
\end{revision}
We estimate completeness as
\begin{equation}
C(J,z)=\frac{N_{\mathrm{det}}(J,z)}{N_{\mathrm{tot}}(J,z)},
\qquad
N_{\mathrm{tot}}(J,z)=5\times10^{4}.
\end{equation}

\section{Results}
\label{sec:results}

\subsection{Detection completeness as a function of magnitude and redshift}
\label{sec:results_qso_completeness}

\begin{figure*}
    \centering
    \includegraphics[width=\linewidth]{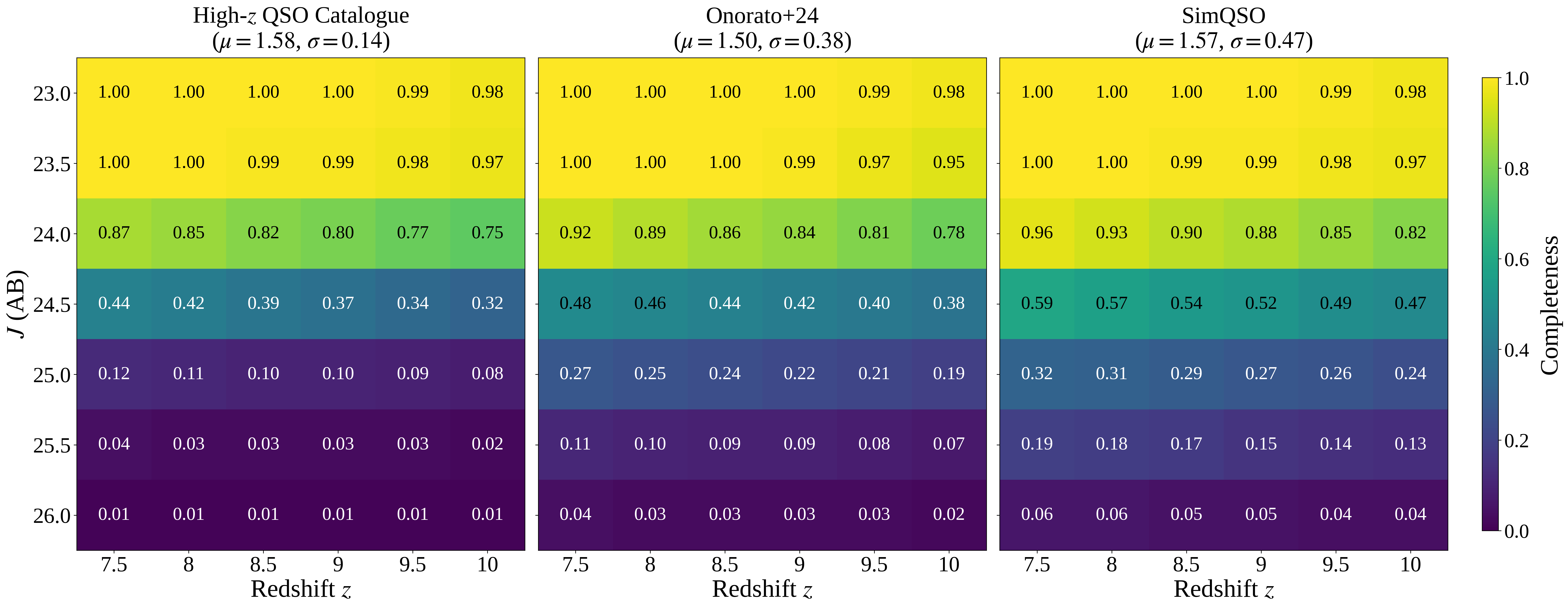}
    \caption{
        Ly$\alpha$-break detection completeness $C(J,z)$ across the
        three Ly$\alpha$ EW priors propagated through the simulation
        pipeline. Each panel shows the fraction of Monte Carlo
        realizations satisfying $\mathrm{SNR}_{\rm break} \ge 3$ on a
        grid of apparent $J$-band magnitude (vertical) and redshift
        (horizontal). \rev{The figure shows the seven completeness bins
        $J_{\rm AB}=23.0$--26.0; the auxiliary $J=26.5$ forward-model bin
        has $C=0$ throughout and is omitted.} \textit{Left:} narrow high-$z$ QSO catalogue
        prior ($\mu = 1.58$, $\sigma = 0.14$ in
        $\log_{10}\mathrm{EW}/\text{\AA}$). \textit{Middle:}
        Onorato-like prior ($\mu = 1.50$, $\sigma = 0.38$).
        \textit{Right:} SimQSO-like prior ($\mu = 1.57$, $\sigma =
        0.47$).
    }
    \label{fig:lya_completeness}
\end{figure*}

\begin{revision}
\newrev{Figure~\ref{fig:lya_completeness} shows break-detection completeness evaluated at the Ly$\alpha$ wavelength set by each simulated input redshift (``redshift-conditioned'' completeness), rather than completeness from a blind search over redshift.}  At $J_{\rm AB}=23.5$ all three give $C\simeq0.95$--1.00. The models separate quickly at fainter magnitudes: at $J=24.5$, $C=0.32$--0.59 across redshift and EW prior, and at $J=25.0$ the range is 0.08--0.32. \newrev{We define $J_{90}$, $J_{50}$, and $J_{10}$ as the apparent magnitudes at which the simulated break-detection completeness $C(J,z)$ falls to 90\%, 50\%, and 10\%, respectively, for a specified redshift and EW prior. Linear interpolation yields $J_{90}=23.65$--24.08, $J_{50}=24.29$--24.67, and $J_{10}=24.96$--25.85 across the simulated redshifts and priors.} In practical terms, the break is recovered with high completeness to about $J\simeq24$, falls through 50\% near $J\simeq24.4$--24.6, and develops a strongly EW-dependent tail beyond $J\simeq25$. \newrev{All three priors fall to $C\le0.06$ by $J_{\rm AB}=26.0$. The surfaces diverge most at $J=25.0$--25.5: averaged over redshift, the Onorato-like and SimQSO-like priors recover 2.3 and 2.8 times as many sources as the narrow prior at $J=25$, and 3.0 and 5.3 times as many at $J=25.5$. The faint-end yield is set by the strength of the high-EW tail of the prior.} For the two broader priors, the drop from $z=7.5$ to 10 is about 20\% at $J=24.5$ and 25--30\% at $J=25$.
\end{revision}

% Completeness tables removed in revision; Figure~\ref{fig:lya_completeness} and the concise text summarize the same information.

\subsection{Spectral-angle classification of detected spectra}
\label{sec:purity}

\begin{revision}
We next apply the spectral-angle discriminant of Section~\ref{sec:sam} to the QSO, brown-dwarf, and passive-galaxy spectra. We use $\Delta=0$ as the nearest-template boundary for this experiment. As noted above, the result depends on the fixed-redshift setup and on the particular reference libraries.
\end{revision}

\begin{figure*}[t]
\centering
\includegraphics[width=\textwidth]{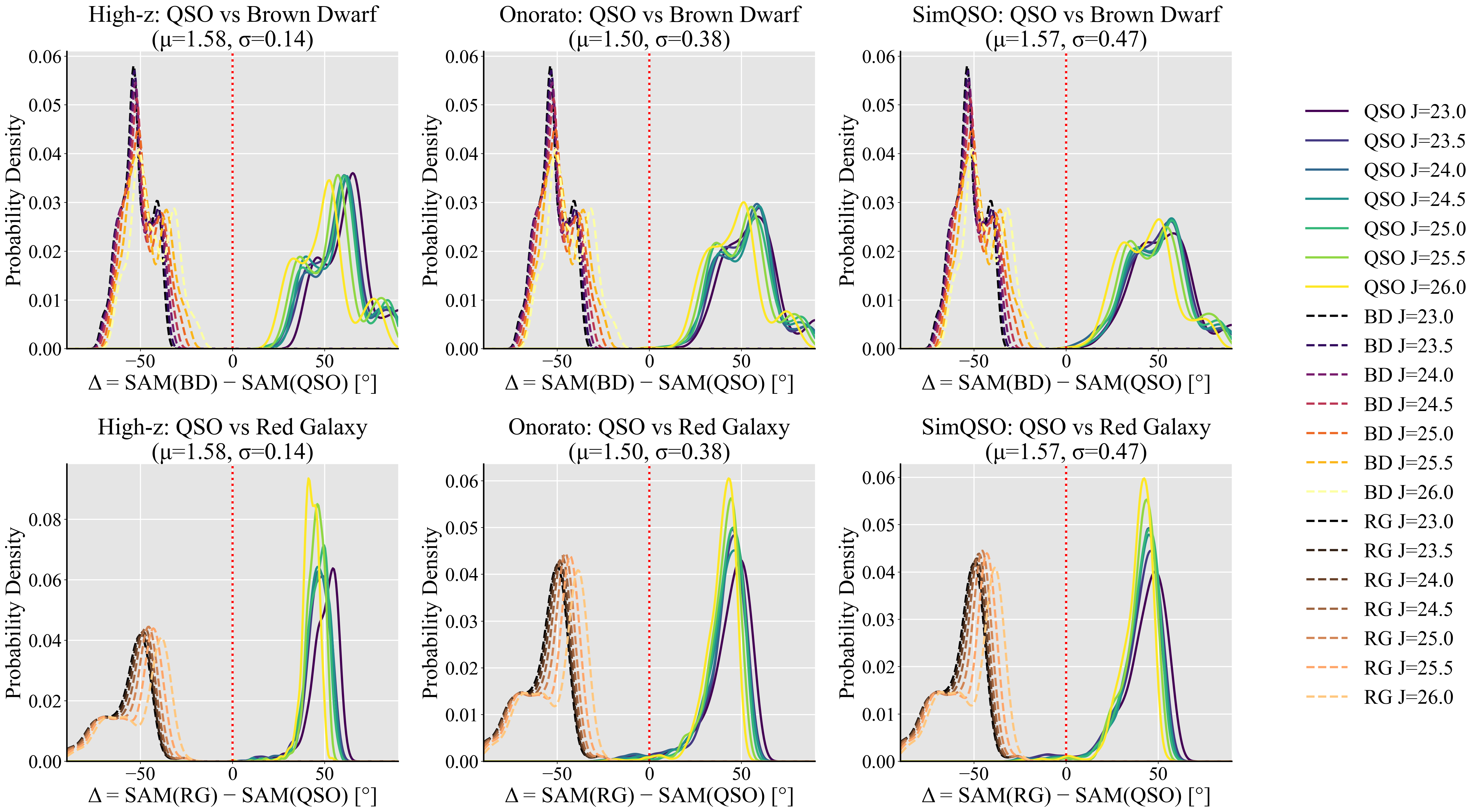}
\caption{Distribution of the spectral-angle discriminant
$\Delta = \mathrm{SAM}_{\mathrm{cont}} - \mathrm{SAM}_{\mathrm{QSO}}$
for QSOs (solid) and contaminants (dashed) as a function of $J$-band
magnitude (displayed bins $J_{\mathrm{AB}} = 23.0$--$26.0$; the stored classifier grid also includes $J=26.5$).
\textit{Top row:} brown-dwarf contaminants.
\textit{Bottom row:} red-galaxy contaminants.
Columns correspond to the three Ly$\alpha$ EW priors (High-z, Onorato,
SimQSO; $\mu$ and $\sigma$ are the mean and scatter of the
$\log_{10}\mathrm{EW}$ distribution in \AA). The red dotted line marks the decision boundary $\Delta=0$. \rev{None of the stored brown-dwarf or red-galaxy trials crosses to the QSO side of the boundary. The QSO recovery rate is $\ge0.982$ across the three EW priors, with a small QSO tail extending to the contaminant side.}}
\label{fig:delta_plot}
\end{figure*}

Figure~\ref{fig:delta_plot} shows the $\Delta$ distributions for QSOs, brown
dwarfs, and red galaxies across the three EW priors and magnitude bins
extending to $J_{\rm AB}=26.0$. \rev{None of the stored contaminant trials crosses to $\Delta>0$, while the QSO recovery rate remains at least 0.982. The QSO and contaminant distributions are therefore well separated, although a small fraction of QSOs falls on the contaminant side.}

\begin{revision}
Table~\ref{tab:sam_results} gives the QSO recovery rates at $\Delta=0$. Against the brown-dwarf library the TPR is 1.000 for all three EW priors; against the red-galaxy library it is 1.000, 0.985, and 0.982 for the high-$z$, Onorato-like, and SimQSO-like cases. None of the stored $N_{\rm BD}=8000$ brown-dwarf or $N_{\rm RG}=11360$ red-galaxy trials crosses the boundary. If those trials were independent binomial draws, zero events would correspond to formal 95\% upper bounds of $3.744\times10^{-4}$ and $2.637\times10^{-4}$. They are not independent parent-object tests, however: the noisy contaminants reuse the reference parent libraries, and some red-galaxy vectors include zero padding outside the native spectral range. We therefore quote the zero-crossing result as a library noise-stability diagnostic. A leave-parent-out run is needed before assigning a survey FPR.
\end{revision}

\begin{figure}[t]
\centering
\includegraphics[width=0.5\textwidth]{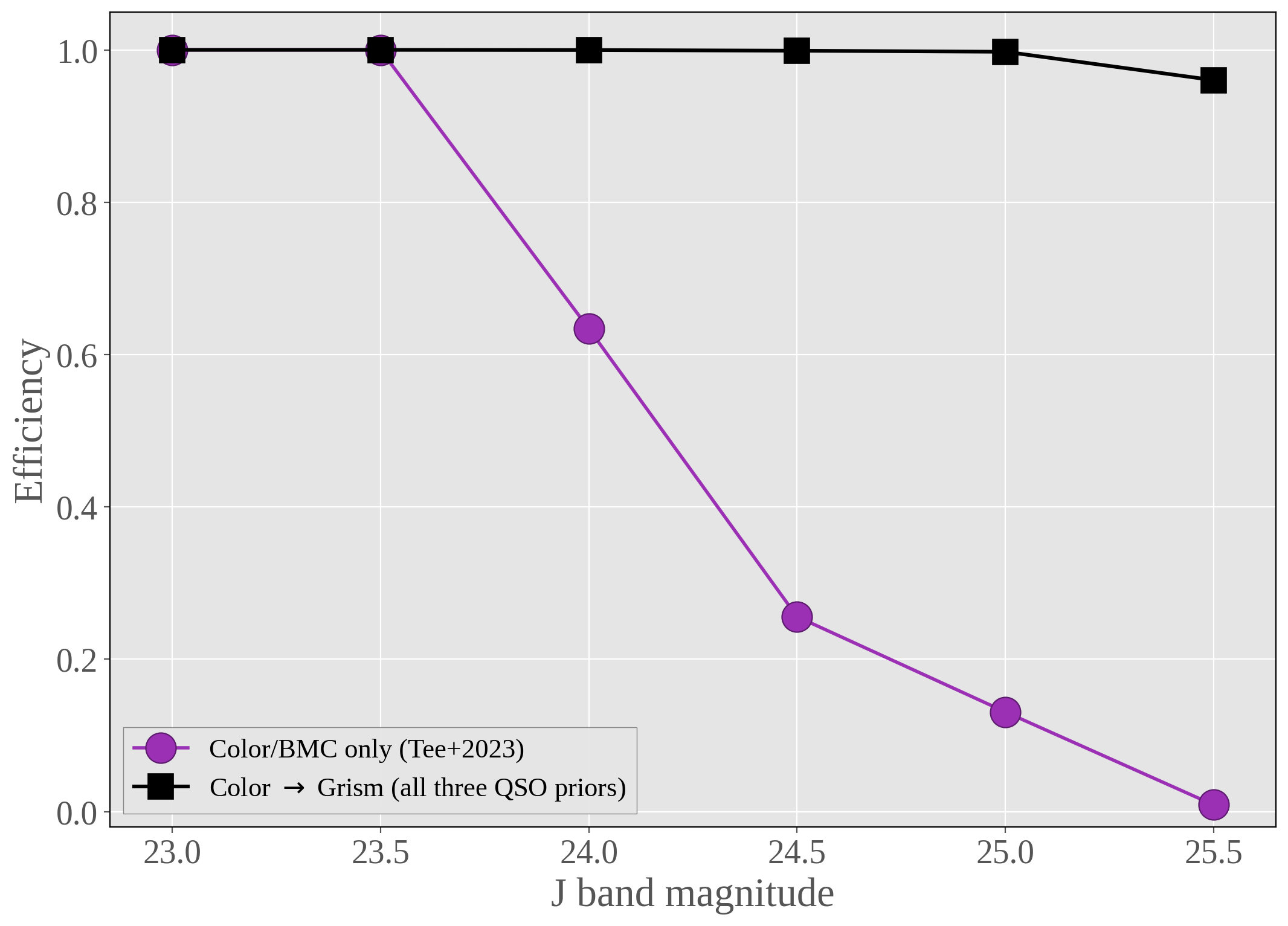}
\caption{
\newrev{Candidate-sample purity $\epsilon=\mathrm{TP}/(\mathrm{TP}+\mathrm{FP})$ versus $J$-band
magnitude. Purple circles show the prior purity $\pi(J)$ of the color-selected sample from}
\citet{Tee2023} (Table~3, Overall column, Color Dropout$+$BMC with
$P_q\ge 0.1$); values at $J\le 23.5$ are at the saturated bright-end
limit, where \citet{Tee2023} report $\pi\to 1$. Black squares show the
posterior efficiency
$\epsilon_{\mathrm{post}}(J)=\mathrm{TPR}\,\pi/(\mathrm{TPR}\,\pi+\mathrm{FPR}(1-\pi))$
after the additional grism (SAM) confirmation step. \rev{For $\mathrm{TPR}=0.982$ and $\mathrm{FPR}=3.744\times10^{-4}$, $\epsilon_{\rm post}=0.99978$, 0.99889, 0.99748, and 0.95971 at $J=24.0$, 24.5, 25.0, and 25.5. The FPR input is the same-parent contaminant bound, so the curve illustrates the color-plus-grism calculation rather than calibrating survey purity.}
}
\label{fig:j_vs_efficiency}
\end{figure}

\begin{revision}
\newrev{For the illustrative ``color-then-grism'' calculation, color/Bayesian photometric selection first generates a candidate list and a grism classification would then be applied to that list. We denote by $\pi(J)=P(\mathrm{QSO}\mid\mathrm{photometric\ candidate},J)$ the fraction of genuine QSOs among the photometrically selected candidates at magnitude $J$, before grism selection. We denote by $\epsilon_{\mathrm{post}}(J)=P(\mathrm{QSO}\mid\mathrm{photometric\ candidate,grism\ pass},J)$ the fraction of genuine QSOs among objects that pass both stages, i.e.\ the purity of the final sample. TPR and FPR are the probabilities that a genuine QSO and a contaminant, respectively, pass the grism step. Bayes' rule then gives}
\begin{equation}
\epsilon_{\mathrm{post}}(J)=
\frac{\mathrm{TPR}\,\pi(J)}
{\mathrm{TPR}\,\pi(J)+\mathrm{FPR}[1-\pi(J)]}.
\label{eq:post_eff}
\end{equation}
\newrev{For $\pi(J)=1$ the equation gives $\epsilon_{\mathrm{post}}=1$ for any $\mathrm{TPR}>0$. This is the intended behavior: a candidate sample that contains no contaminants stays pure after a step that can only remove objects. The grism step then costs completeness, a fraction $1-\mathrm{TPR}$ of the genuine QSOs, while the purity is unchanged. Inserting $3.744\times10^{-4}$ as the FPR, the lowest measured recovery rate $\mathrm{TPR}=0.982$, and $\pi(J)=\{0.634,0.256,0.131,0.009\}$ from \citet{Tee2023} gives $\epsilon_{\mathrm{post}}=0.999780$, 0.998893, 0.997477, and 0.959711 at $J=24.0$, 24.5, 25.0, and 25.5. At $J=25.5$ the FPR term is 4.20\% of the TPR term. Because the FPR input comes from the same-parent contaminant test, these values illustrate the color-plus-grism calculation rather than calibrate survey purity. Figure~\ref{fig:j_vs_efficiency} shows the resulting curve.}
\end{revision}

\begin{table}[h]
\centering
\caption{\rev{QSO recovery at $\Delta=0$ in the fixed-redshift SAM test. The columns give the QSO TPR against each contaminant library. None of the stored noisy contaminant trials crosses the boundary. Because the noisy tests reuse the parent reference libraries, the zero-crossing result measures noise stability within those libraries rather than an independent survey FPR.}}
\label{tab:sam_results}
\begin{tabular}{lcc}
\hline\hline
EW prior & $\mathrm{TPR}_{\mathrm{QSO|BD}}$ & $\mathrm{TPR}_{\mathrm{QSO|RG}}$ \\
\hline
High-$z$ ($\sigma=0.14$ in $\log_{10}\mathrm{EW}$)  & 1.000 & 1.000 \\
Onorato-like ($\sigma=0.38$) & 1.000 & 0.985 \\
SimQSO-like ($\sigma=0.47$)  & 1.000 & 0.982 \\
\hline
\end{tabular}
\end{table}

\begin{revision}
The completeness and SAM experiments use the input redshift in different ways. The break statistic is measured at the expected Ly$\alpha$ wavelength, and the QSO SAM comparison is restricted to references at that redshift. These tests therefore address detectability and spectral-shape separation at known $z$; they do not measure the depth of a blind redshift search. That would require scanning redshift, accounting for the additional trials, and testing spectra outside the present QSO model family.
\end{revision}

\section{Discussion}

\subsection{The selection function of Roman grism spectroscopy at the faint end}
\label{sec:discussion_faintest_qso}

The completeness analysis presented in Section~\ref{sec:results} demonstrates
that Roman's faint-QSO reach is not a single instrumental magnitude limit.
Multiple factors contribute to whether a $z>7.5$ QSO is recovered: the
apparent $J$-band magnitude, the source redshift (which sets the IGM
transmission and the location of the Ly$\alpha$ break in the grism bandpass),
and the intrinsic Ly$\alpha$ equivalent width drawn from the relevant
population distribution. Here we discuss what this implies for the limiting
depth of a Roman wide-area QSO survey, drawing on the $C(J,z)$ surfaces of
Section~\ref{sec:results_qso_completeness}.

The behavior of $C(J,z)$ separates into three magnitude regimes. At
$J_{\rm AB}\le 23.5$, the red-side continuum is bright enough that the break
test is dominated by continuum contrast, and all three EW priors give
$C\simeq 0.95$--$1.00$. In this regime the uncertainty in the Ly$\alpha$ EW
distribution does not propagate into the recovery rate. At intermediate
magnitudes, $J_{\rm AB}=24.0$--$24.5$, the priors begin to separate, with the
high-$z$ prior dropping from $C\simeq 0.87$--$0.75$ at $J=24.0$ to
$C\simeq 0.44$--$0.32$ at $J=24.5$, and the broader Onorato and SimQSO priors
retaining $C\simeq 0.48$--$0.38$ and $C\simeq 0.59$--$0.47$ respectively. At
$J_{\rm AB}=25.0$ the prior dependence is largest: the high-$z$ prior gives
$C\simeq 0.12$--$0.08$, while the Onorato and SimQSO priors give
$C\simeq 0.27$--$0.19$ and $C\simeq 0.32$--$0.24$, a factor of $\sim 2$--$3$
higher than the narrow prior at fixed $(J,z)$. The faint-end reach therefore
depends on which Ly$\alpha$ EW distribution is assumed to apply at
$z\gtrsim 7.5$, a question that remains observationally unsettled given the
limited spectroscopic samples available at these redshifts \citep{Yang2021ApJ}.

There are also strong reasons to expect that the EW prior dependence reported
here may underestimate the true uncertainty in faint-end yield. The three
priors used in this work are calibrated against samples that lie almost
entirely at $z\lesssim 6$, where the population of bright QSOs has been
characterized through optical and near-infrared spectroscopy
\citep{Yang2021ApJ,Onorato2025,SimQSO2021}. Whether the Ly$\alpha$ EW distribution
at $z\gtrsim 7.5$ is narrower, broader, or shifted relative to the
$z\lesssim 6$ samples is not directly constrained, since the few QSOs known
at $z>6.5$ are by selection biased toward bright luminosities and toward
detectable Ly$\alpha$. A measurement of the EW distribution at $z>7.5$ would
in turn require a relatively unbiased spectroscopic sample at the magnitudes
where Roman's selection function changes most rapidly, which is precisely the
regime that Roman would itself probe. The three priors used here should
therefore be interpreted as a bracketing of the dominant astrophysical
uncertainty, not as a calibrated probability distribution over outcomes.

A second factor that is worth examining more carefully is the redshift
dependence of $C(J,z)$ within each prior. At fixed $J$, completeness
decreases monotonically from $z=7.5$ to $z=10$, but the decrease is gradual:
\rev{a relative drop of about 20\% at $J_{\rm AB}=24.5$ and 25--30\% at $J_{\rm AB}=25$ for the broader priors. Blueward of Ly$\alpha$, the transmitted flux is already close to zero throughout this redshift range; the redshift-dependent IGM term in our model is mainly the damping wing immediately redward of the line. At the same time, Ly$\alpha$ moves from $1.033$ to $1.337\,\mu$m, toward the broad maximum of the Roman grism effective area near $1.4\,\mu$m \citep{RomanWFI2025}. Throughput alone would therefore make the break somewhat easier, not harder, to detect at the high-$z$ end. \newrev{The decline in $C(J,z)$ comes from the source and IGM model together with the fixed-band normalization (Section~\ref{sec:sed}), with the throughput modifying the size of the effect.}}

It is useful to translate the magnitude regime accessed by Roman into
intrinsic luminosity. At $z\simeq 8$, $J_{\rm AB}\simeq 24.5$--$25$
corresponds approximately to $M_{1450}\simeq -22.5$ to $-22.0$, or
$L_{\rm bol}\sim 10^{45}\,{\rm erg\,s^{-1}}$ for standard UV bolometric
corrections. This is fainter than the classical wide-field $z>6$ QSOs
\citep{banados2018,Wang2021QSO}, where $M_{1450}\lesssim -25$, and brighter than
many of the broad-line AGN candidates now being identified in deep
\textit{JWST} fields \citep{Harikane2023,Maiolino2023}, where $L_{\rm bol}$
ranges from $\sim 10^{43}$ to $\sim 10^{45}\,{\rm erg\,s^{-1}}$. Roman's
forecast reach therefore lies in an intermediate-luminosity regime that has
been only sparsely sampled at $z>7.5$ to date.

\begin{revision}
The SAM separation is strong for the two contaminant classes we actually tested, M/L/T dwarfs and passive red galaxies: the QSO TPR is at least 0.982 and none of the stored contaminant trials crosses $\Delta=0$. The latter result should be read with care. The contaminant tests reuse the parent reference libraries, and some red-galaxy vectors contain zero-padded wavelengths that tend to push $\Delta$ farther onto the contaminant side, so this is not an independent FPR measurement. We also did not include star-forming galaxies at $z=7.5$--10. Such galaxies can show the same Lyman break and narrow Ly$\alpha$ emission as a quasar candidate. Roman can help when a broad component is detected: at the observed Ly$\alpha$ wavelengths the point-source resolution is $R\simeq475$--620 (about 630--490~km~s$^{-1}$), so a several-thousand km~s$^{-1}$ line spans several resolution elements; \ion{C}{4} also remains in the grism band. At low S/N, or when the broad component is weak, the high-$z$ galaxy degeneracy remains. The QSO TPR has a separate limitation: both reference and test spectra come from the same broad spectral model family, so it is an in-family result.
\end{revision}

\begin{revision}
For the completeness calculation, the main astrophysical sensitivity is the adopted Ly$\alpha$ EW distribution, especially the strong-line tail. The SAM test shows that the simulated QSOs separate well from the present brown-dwarf and passive-galaxy libraries at fixed redshift. It does not establish near-unity purity for a real Roman catalogue; that requires broader and independent QSO and contaminant samples.
\end{revision}

\subsection{Limitations of the classification forecast}
\label{sec:classification_limitations}
\begin{revision}
Several limitations matter for the classification result. We compare QSOs at the known input redshift, and the QSO reference and test spectra share the same underlying spectral model. The contaminant set contains only M/L/T dwarfs and passive red galaxies. In addition, the contaminant trials are not leave-parent-out, and some red-galaxy vectors include zero padding where the parent spectrum has no coverage. We have not modeled high-$z$ star-forming galaxies, trace overlap, or a blind redshift search. For these reasons, the reported TPR and zero-crossing counts describe the present simulation setup rather than the final purity of a Roman survey sample.
\end{revision}

\subsection{Roman in the context of \textit{JWST}, Euclid, and Rubin/LSST}
\label{sec:discussion_multi_survey}

\rev{The $J_{\rm AB}\simeq24.5$--25 regime ($M_{1450}\simeq-22.5$ to $-22.0$) overlaps the faint end already reached by SHELLQs at $z\sim6$, while extending the comparison toward the higher-redshift population probed by Roman and the much fainter AGN found in deep \textit{JWST} fields.} This is where the Roman completeness
surface $C(J,z)$ begins to transition from high completeness to
prior-dependent incompleteness, where broad-band selections become more
sensitive to photometric scatter and contaminant overlap, and where the
\textit{JWST} faint-AGN population begins to connect to the brighter
wide-field quasar samples. A Roman-confirmed sample in this regime could
therefore test whether the QSO luminosity function connects smoothly
between the classical wide-area bright end and the deep-field faint-AGN
population, or whether the transition includes a break, a duty-cycle change,
or an evolving fraction of objects with detectable Ly$\alpha$.
Such a measurement would test whether the potentially substantial,
model-dependent AGN contribution inferred at $z\sim5$--6
\citep{Singha2025Ionizing} persists earlier in reionization.
We do not forecast that luminosity function here. \newrev{We provide a first estimate of two of its ingredients: the known-redshift break completeness $C(J,z)$ for isolated traces (Section~\ref{sec:results_qso_completeness}) and the separation of simulated QSOs from the tested contaminant libraries (Section~\ref{sec:purity}). Both are model-dependent and will need to be recalibrated on Roman data, where trace overlap, blind redshift search, and contaminant classes we have not simulated enter.}
\section*{Conclusions}

We investigated Roman's potential to detect faint $z>7.5$ quasars and
separate them from cool stellar and red-galaxy contaminants in HLWAS-like
slitless grism observations. We built an end-to-end forward model that
generates synthetic QSO SEDs under three Ly$\alpha$ equivalent-width priors
(narrow high-$z$, Onorato-like, SimQSO-like) and applies an IGM
transmission model with patchy reionization. We propagated each spectrum
through PSF-convolved grism dispersion onto the WFI detector and extracted
1D spectra via Horne-style optimal extraction. We measured Ly$\alpha$-break
detection completeness on a $(J_\mathrm{AB}, z)$ grid under each prior. We
classified the extracted spectra against M, L, and T dwarf and passive
red-galaxy template libraries using the spectral-angle mapper. \rev{The Ly$\alpha$ break is recovered with high completeness to roughly $J_{\rm AB}\simeq24$, with $C=0.5$ near $J_{\rm AB}\simeq24.4$--24.6 and an EW-dependent tail beyond $J_{\rm AB}\simeq25$. At fixed input redshift, the QSO recovery rate against the present M/L/T-dwarf and passive-galaxy libraries is at least 0.982. These classification numbers describe the tested libraries and should not be read as a survey-wide purity measurement.} This places Roman in the intermediate-luminosity regime between
classical wide-field QSO surveys and the faint AGN population emerging from
\textit{JWST}, allowing us to test whether the QSO luminosity function
continues smoothly to faint magnitudes and, with an empirically calibrated selection function, to improve constraints on the AGN contribution to reionization.

\begin{acknowledgments}
%\textit{Acknowledgements.}
The material is based upon work supported by NASA under award number 80GSFC21M0002. 

\end{acknowledgments}

%%%%%%%%%%%% Table of Radio Quiet unabsorbed Fluxes

%%%%%%%%%%%%%%%%%%%%%%%%%%%%%%%%%%%%%%%%%%%%%%%%%%%%
\newpage

\bibliography{bib1}
\pagebreak

\end{document}